\documentclass[doublecol]{epl2} 

\title{Coexistence of long-ranged charge and orbital order and spin-glass state in single-layered manganites with weak quenched disorder}
\shorttitle{Charge and orbital order and spin-glass state in single-layered manganites}

\author{R. Mathieu\inst{*,1}, J. P. He\inst{1}, X. Z. Yu\inst{1}, Y. Kaneko\inst{1,2}, M. Uchida\inst{1}, Y. S. Lee\inst{1}, T. Arima\inst{1,3}, A. Asamitsu\inst{1,4}, Y. Tokura\inst{1,2,5,6}}
\shortauthor{R. Mathieu \etal}

\institute{
  \inst{*}  Now at:  Department of Microelectronics and Applied Physics, Condensed Matter Physics group (KMF), Royal Institute of Technology (KTH), Electrum 229, SE-164 40 Kista, Sweden. Electronic address: rmathieu@kth.se\\
  \inst{1} Spin Superstructure Project (ERATO-SSS), JST, AIST Central 4, Tsukuba 305-8562, Japan\\
  \inst{2} ERATO multiferroics project, JST, c/o Department of Applied Physics, University of Tokyo, Tokyo 113-8656, Japan\\
  \inst{3} Institute of Multidisciplinary Research for Advanced Materials, Tohoku University, Sendai 980-8577, Japan\\
  \inst{4} Cryogenic Research Center (CRC), University of Tokyo, Bunkyo-ku, Tokyo 113-0032, Japan\\
  \inst{5} Correlated Electron Research Center (CERC), AIST Central 4, Tsukuba 305-8562, Japan\\
  \inst{6} Department of Applied Physics, University of Tokyo, Tokyo 113-8656, Japan
}


\abstract{
The relationship between orbital and spin degrees of freedom in the single-crystals of the hole-doped Pr$_{1-x}$Ca$_{1+x}$MnO$_4$, 0.3 $\leq$ $x$ $\leq$ 0.7, has been investigated by means of ac-magnetometry and charge transport. We show that in an intermediate underdoped region, with 0.35 $\leq$ $x$ $<$ 0.5, the ``orbital-master spin-slave'' relationship commonly observed in half-doped manganites does not take place. The long-ranged charge-orbital order is not accompanied by an antiferromagnetic transition at low temperatures, but by a frustrated short-ranged magnetic state bringing forth a spin-glass phase. We discuss in detail the nature and origin of this true spin-glass state, which, as in the half-doped manganites with large quenched disorder, is not related to the macroscopic phase separation observed in crystals with minor defects or impurities.
}

\usepackage{color}
\newcommand{\revision}[1]{{\color{black}{#1}}}

\begin{document}

\maketitle

\begin{figure}[h]
\onefigure[width=0.46\textwidth]{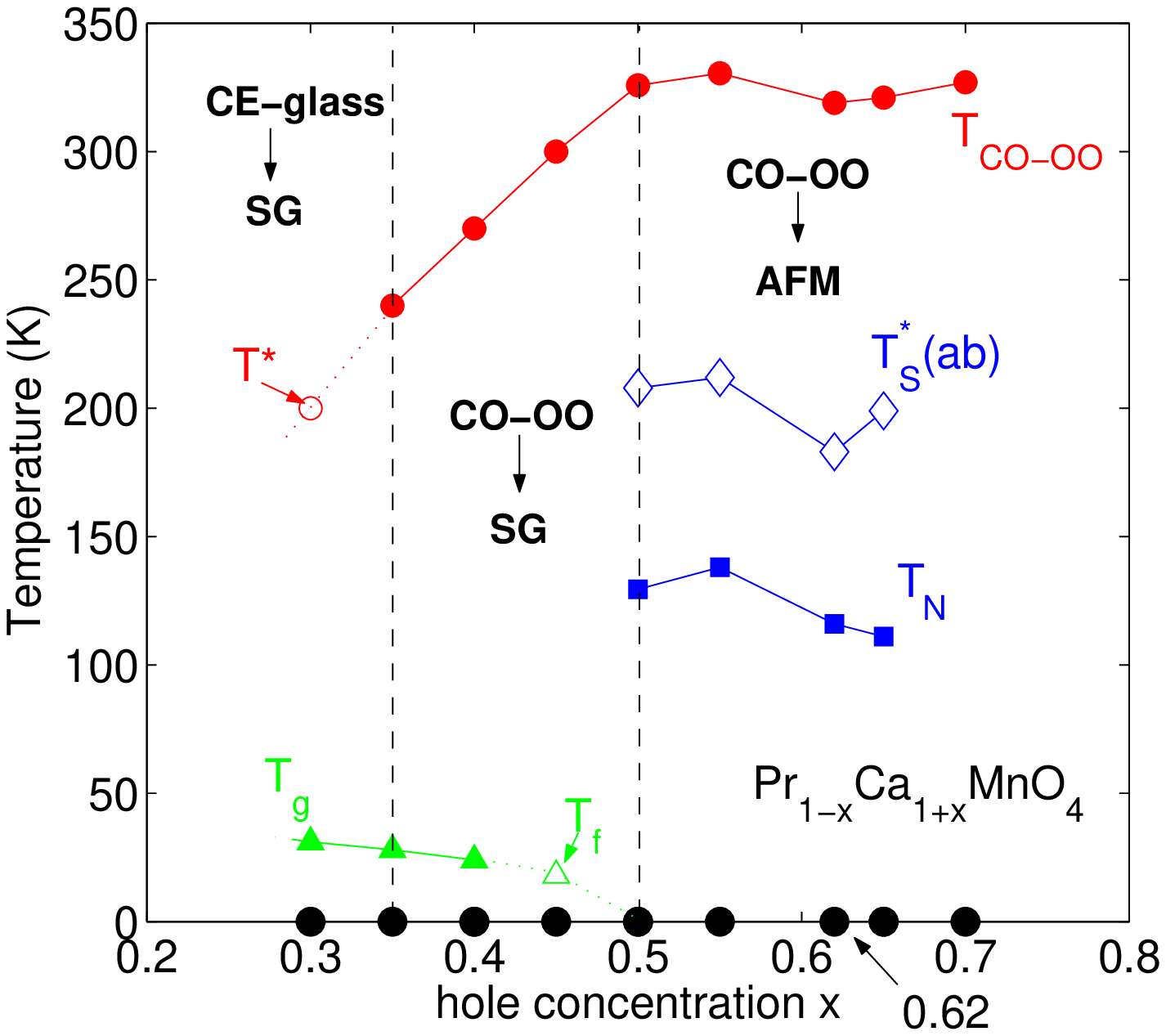}
\onefigure[width=0.46\textwidth]{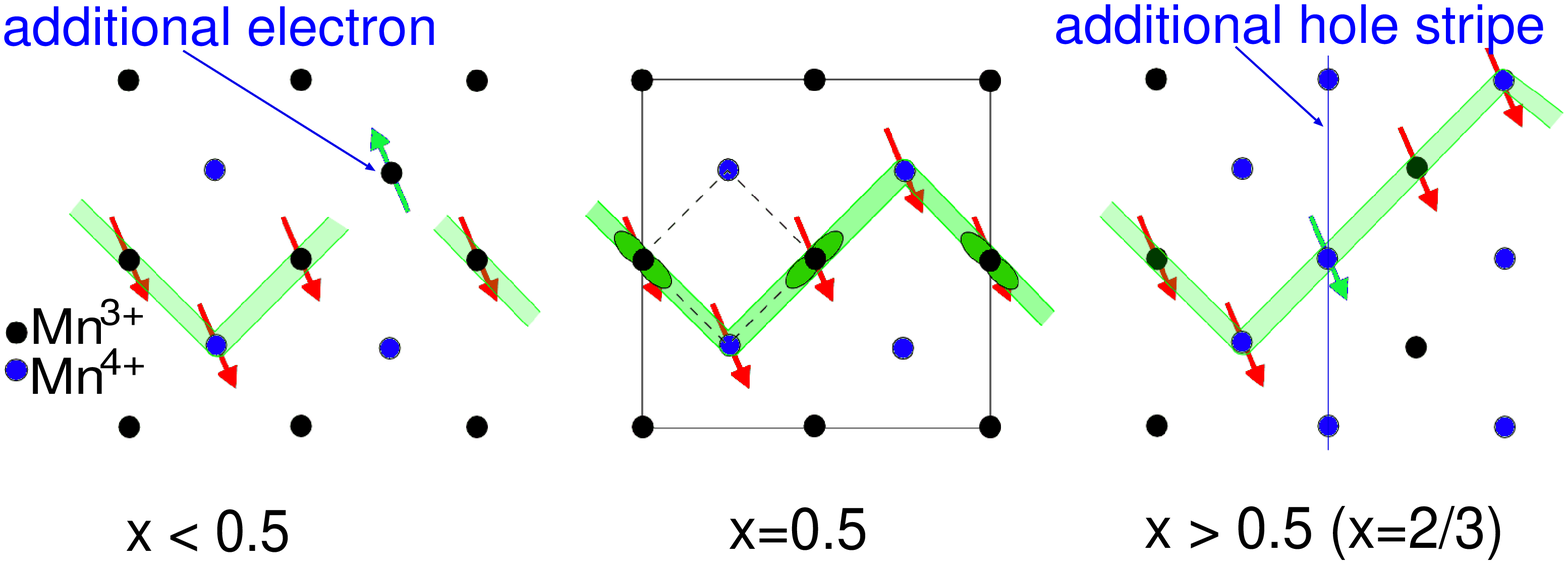}
\caption{Upper panel: Electronic phase diagram for all crystals. The open symbols mark the onsets of the short-ranged orbital ($T^*$, estimated from the ED; see Ref.~\cite{yu}) and magnetic ($T_{\rm S}^*(ab)$) correlation, and of the non-equilibrium dynamics ($T_f$). Lower panels: schematic representation of the charge- and orbital ordered states of underdoped ($x<$0.5), half-doped ($x$=0.5) and overdoped  ($x>$0.5) crystals, showing the magnetic arrangement within the zig-zag chains of the CE-type magnetic structure. The orbital order involves staggered $3x^2-r^2/3y^2-r^2$ orbitals of the $e_g$-like electrons of Mn$^{3+}$, represented as green (dark gray) lobes in the figure. The spins, represented with red (dark gray) arrows, order ferromagnetically along zig-zag chains, a fragment of which is highlighted in light green (light gray) in the figure. The squares in the $x$=0.5 panel represent the magnetic (continuous line) and $I4/mmm$ (dashed line) unit cells respectively.}
\label{fig-diag}
\end{figure}

The strong coupling among spin, charge, orbital, and lattice degrees of freedom\cite{orbital} is a key ingredient for the colossal magnetoresistive (CMR) properties of manganese oxides. For example, the observation of the CMR effect in hole-doped $R_{1-x}A_x$MnO$_3$ manganites ($R$ and $A$ being rare earth and alkaline earth ions, respectively) is not only determined by the one-electron bandwidth $W$, but also by the degree of quenched disorder\cite{aka}. Depending on the radii of the $R$ and $A$ cations, the ferromagnetic (FM) metallic phase, or the charge- and orbital-ordered (CO-OO) phase can be stabilised\cite{tomioka-diag}. Phase diagrams in the plane of $r_A$ (the average ionic radius, related to $W$ ) and $\sigma^2$ (the ionic radius variance, measuring the disorder), or bandwidth-disorder phase diagrams, have been established for nearly half-doped perovskite manganites\cite{tomioka-diag}, with a three-dimensional (3$D$) $Mn-O$ network, as well as half-doped single-layered manganites with a two-dimensional (2$D$) $Mn-O$ network\cite{roland-diag}. Considering only the low-bandwidth region for the 3$D$ perovskites, both phase diagrams only contain two regions, with long-ranged and short-ranged CO-OO order respectively\cite{tomioka-diag, roland-diag}. An ``orbital-master spin-slave'' relationship exists between these two degrees of freedom for the manganites with strong Jahn-Teller coupling. The long-ranged CO-OO state, usually described in terms of staggered 3$x^2-r^2$ and 3$y^2-r^2$ orbitals, is accompanied by a long-ranged CE-type antiferromagnetic (AFM) structure, composed of ferromagnetic zig-zag chains antiferromagnetically coupled with each other in the $ab$-plane\cite{jirak} (see the lower panels of Fig.~\ref{fig-diag} for a schematic view of the orbital-spin arrangement). The short-ranged CO-OO state, or CE-glass state\cite{dagotto,roland-diag}, is instead accompanied by a short-ranged spin-glass (SG) state\cite{roland-esmo,roland-ebmo}. In 3$D$, the CO-OO collapses in a first-order manner, while in 2$D$ the disappearance of the CO-OO state is more gradual\cite{roland-diag,uchida}. The observed SG state is associated with the glassy state of the orbital sector rather than with magnetic inhomogeneity or micrometer-scale phase separation\cite{roland-esmo,roland-ebmo}. In any case, the phenomenon of  macroscopic phase separation\cite{raveau,kimura} does not occur\cite{tomioka-diag,roland-diag}.\\

In the present letter, we have investigated the magnetic and electrical properties of high-quality single crystals of the single layered Pr$_{1-x}$Ca$_{1+x}$MnO$_4$, (0.3 $\leq$ $x$ $\leq$ 0.7) with weak quenched disorder. Even though there is no cation ordering on the $A$-site, the quenched disorder is extremely weak in this system due to the very similar ionic size of  Pr$^{3+}$ and Ca$^{2+}$. We have observed  the asymmetric response of the orbital and spin sectors of the system to excess/deficiency of holes with respect to the half-doping case ($x$=0.5). The long-ranged CO-OO state observed for $x$=0.5 is rearranged to accommodate additional holes (over-doping, $x$ $>$ 0.5), while it is weakened by the presence of additional electrons (under-doping, $x$ $<$ 0.5) and eventually vanishes at $x$=0.3. In the  0.35 $\leq$ $x$ $<$ 0.5 range, the extra localised electrons introduce some frustration in the magnetic structure, ultimately yielding the formation of a SG state at low temperatures, while the CO-OO remains long-ranged. Hence interestingly, our results imply that the relationship of spin and orbital degrees of freedom is different than that of the half-doped case. We discuss in detail this relationship, which is not related to the phenomenon of macroscopic phase separation, in the light of recent electron diffraction studies.\\

High quality single crystals of the $A$-site solid-solution Pr$_{1-x}$Ca$_{1+x}$MnO$_4$ (PCMO) manganites were grown by the floating zone method, with the hole doping level $x$ varying from $x$=0.3 to 0.7 in steps of 0.05. The quenched disorder inherent to the solid solution of Pr$^{3+}$/Ca$^{2+}$ ions on the $A$-sites is kept minimum, as both ions are very similar in size. The degree of disorder does not change significantly with $x$, with $\sigma^2$ only varying from 2.27$\times$10$^{-7}${\AA}$^2$ to 1.27$\times$10$^{-7}${\AA}$^2$ as $x$ varies from 0.3 to 0.7. 
\begin{figure}[h]
\onefigure[width=0.46\textwidth]{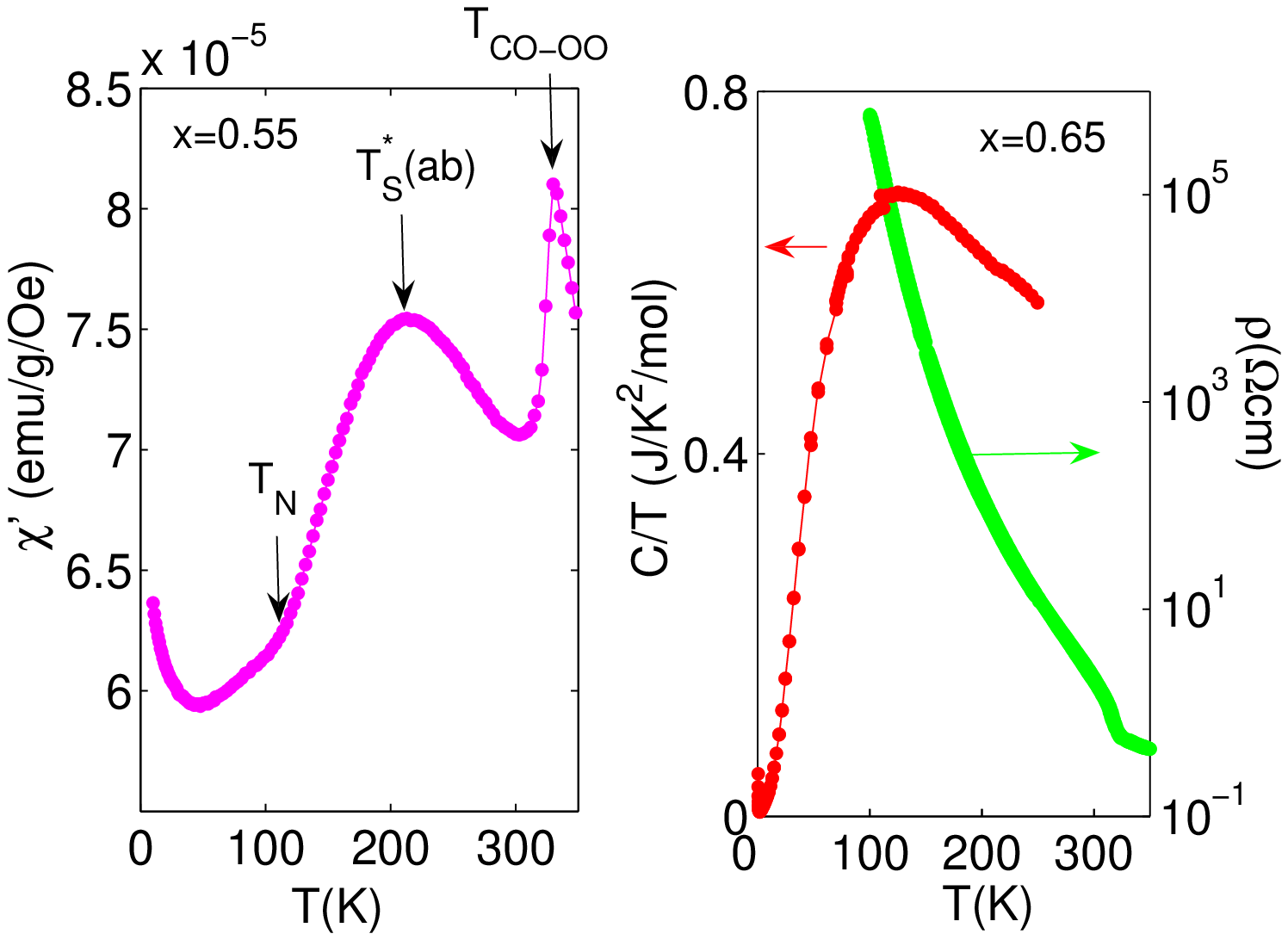}
\caption{\revision{Left: Temperature dependence of the in-phase component of the ac-susceptibility $\chi'$  $\chi'$ for the crystal with $x$=0.55, identifying $T_{CO-OO}$, $T_S^*(ab)$, and $T_N$ (see main text). Right: Temperature dependence of the heat capacity $C$ (as $C/T$) and electrical resistivity $\rho$ for the crystal with $x$=0.65.}}
\label{fig-HCR}
\end{figure}
\begin{figure}[h]
\onefigure[width=0.46\textwidth]{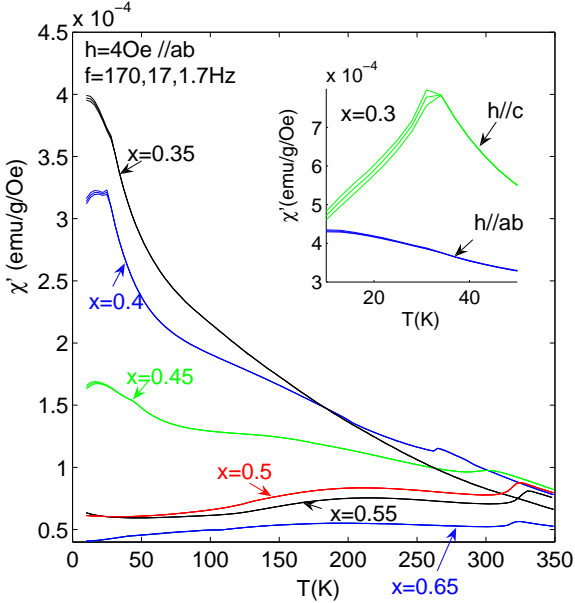}
\caption{\revision{Temperature dependence of the in-phase component of the ac-susceptibility $\chi'$ for selected crystals. The ac field is applied in the $ab$-plane with $h$ (field amplitude) = 4 Oe and $f$  (frequency) = 170, 17, and 1.7 Hz. The inset shows the $T$- and $f$-dependence of the out-of-phase component $\chi''$ for $x$=0.3 with the ac field applied in and perpendicular to the $ab$-plane.}}
\label{fig-XT}
\end{figure}
In comparison, $\sigma^2$ amounts to 1.65$\times$10$^{-3}${\AA}$^2$ in the well known La$_{0.5}$Sr$_{1.5}$MnO$_4$ showing the long-ranged CO-OO\cite{uchida}, and  6.77$\times$10$^{-3}${\AA}$^2$ in the Eu$_{0.5}$Sr$_{1.5}$MnO$_4$ XY spin glass\cite{roland-esmo,uchida}. The phase-purity of the crystals was checked by x-ray diffraction and the cation concentrations of the crystals was confirmed by inductively coupled plasma (ICP) spectroscopy. The ac-susceptibility $\chi$($\omega=2\pi f$) data was recorded as a function of the temperature $T$ and frequency $f$ on a MPMSXL SQUID magnetometer equipped with the ultra low-field option (low frequencies) and a PPMS6000 (higher frequencies), after carefully zeroing or compensating the background magnetic fields of the systems. Additional phase corrections were performed for some frequencies. The resistivity $\rho$ of the crystals was measured using a standard four-probe method on a PPMS6000, feeding the electrical current in the $ab$-plane. The heat capacity $C$ was recorded using a relaxation method on the same measurement system.\\

\revision{The half-doped single-layered manganites with small quenched disorder such as Pr$_{1-x}$Ca$_{1+x}$MnO$_4$ undergo a CO-OO phase transition above room temperature, accompanied by an antiferromagnetic phase transition at lower temperatures. Overdoped  Pr$_{1-x}$Ca$_{1+x}$MnO$_4$ crystals, with $x$ $>$ 0.5, show similar magnetic and orbital states. As in the half-doped case\cite{roland-diag}, the ac-susceptibility $\chi'$ of the overdoped PCMO crystals exhibits a sharp peak arising from the quenching of the FM spin fluctuation below $T_{\rm CO-OO}$ $\approx$ 310-330K depending on $x$} (see left panel of Fig.~\ref{fig-HCR}). At lower temperatures, near 200K, a broader peak is observed. This broader peak does not correspond to $T_{\rm N}$ for long-range AFM spin order, which is $\sim$ 120-130K in these crystals\cite{jing}. An inflection, more clearly seen in the $T$-derivative of $\chi'(T)$ (see below), can been seen in the vicinity of these temperatures, and is thus identified as $T_{\rm N}$. The high-temperature broad peak likely stems from the short-range AFM correlation developing in the basal planes of the structure, and is labelled $T_{\rm S}^*(ab)$\cite{roland-diag}. Electron diffraction (ED) studies in Ref.~\cite{roland-diag} indicate a gradual atomic displacement in the 160-220K range, which has also been reported in polycrystalline PCMO samples\cite{raveau2}. However, it is unlikely that $T_{\rm S}^*(ab)$ is related to this gradual structural rearrangement as $T_{\rm S}^*(ab)$ is rather independent of the size of the $R$ cation in $R_{0.5}$Ca$_{1.5}$MnO$_4$\cite{roland-diag}, while the structural rearrangement occurs at higher temperatures when the ionic radius of $R$ decreases (e.g. in the 250-290K temperature range for $R$=Sm; see also Ref.~\cite{yu}). The temperature dependence of the heat capacity and resistivity \revision{of the half- and overdoped crystals is similar to the one depicted in the right panel of Fig.~\ref{fig-HCR} for $x$=0.65}. No clear feature is observed in the vicinity of $T_{\rm N}$, nor $T_{\rm S}^*(ab)$, while the characteristic inflection in the resistivity curves occurs near $T_{\rm CO-OO}$\cite{yu}. In half-doped perovskite manganites with small bandwidth, such as $R_{0.5}$Ca$_{0.5}$MnO$_3$ ($R$=Pr, Sm, etc...), the CO-OO order and the resulting AFM ordering also occur at different temperatures. However, the AFM transition temperature is more easily identified in that case\cite{tomioka-physb}.\\

In the underdoped crystals, the $T_{\rm CO-OO}$ peak in the  $\chi'-T$ curves becomes broader and broader as $x$ decreases, until it vanishes for $x$ $\leq$ 0.35, \revision{as shown in Fig.~\ref{fig-XT}}. As seen in that figure, the susceptibility globally increases as $x$ decreases, and a SG-like frequency dependent cusp appears at low temperatures. Below $x$=0.4, the anomalies in the magnetisation are broader, and it is difficult to assert the long- or short-range character of the CO-OO. The susceptibility curves of the crystal with $x$=0.3 are essentially featureless down to 50K.  \revision{The susceptibility curves recorded with an out-of-plane magnetic field display a similar temperature dependence, except that the SG-like cusp becomes stronger as $x$ decreases. As seen in the inset of Fig.~\ref{fig-XT}, the SG-like cusp is essentially only seen in the $c$-direction for $x$ = 0.3}. Although no anomaly is discernible in the susceptibility curves, ED patterns and real-space images reveal the long-ranged CO-OO for all underdoped crystals with $x$ $\geq$ 0.35, and the short-ranged CO-OO of $x$=0.3\cite{yu}.

\begin{figure}[h]
\onefigure[width=0.46\textwidth]{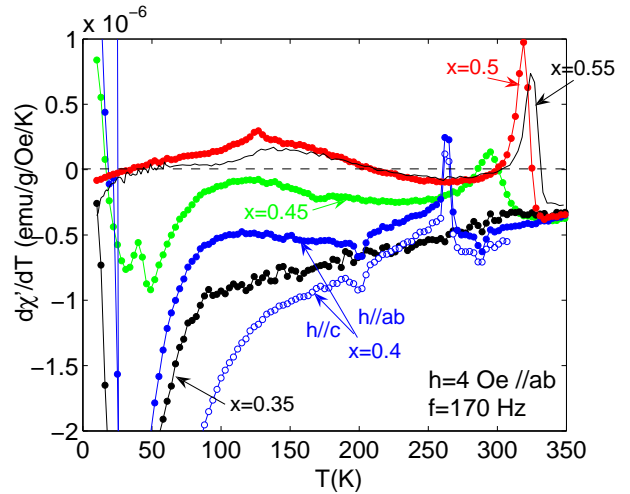}
\caption{\revision{Temperature dependence of the temperature-derivative $d/dT$ of the in-phase component of the ac-susceptibility $\chi'$ for the half- and underdoped crystals. The data for $x$ = 0.55, representative of the overdoped crystals, is included (continuous line) for reference. The ac field was applied in the $ab$ plane. Data for the ac field perpendicular to the $ab$ plane is added in open symbols for comparison for $x$ = 0.4.}}
\label{fig-dXdT}
\end{figure}

In order to investigate in more detail the magnetic properties of the underdoped crystals, we have plotted in Fig.~\ref{fig-dXdT} the temperature derivative of $\chi'$ for the underdoped crystals, $d\chi'/dT$. For $x$=0.5 and 0.55, the above mentioned features of the susceptibility are clearly evidenced; two maxima in $\chi'$ or peaks corresponding to $d\chi'/dT$=0 near 325-330K and 200K, respectively. While the slope of  $d\chi'/dT(T)$ at the former extremum is very large (defining a sharp $T_{\rm CO-OO}$ peak), it is moderate at the latter one, as only a broad peak is observed near $T_{\rm S}^*(ab)$. The tiny inflection observed in the $\chi'-T$ curves is more clearly observed in the  $d\chi'/dT$ curves, which accordingly show a maximum near 120K (see Fig.~\ref{fig-dXdT}). As discussed above, we consider that this inflection marks $T_{\rm N}$. While the sharp $T_{\rm CO-OO}$ peak is observed down to $x$=0.4, the broad $T_{\rm S}^*(ab)$ peak and the $T_{\rm N}$ inflection are not observed below $x$=0.5, which suggests that while the CO-OO is still long-ranged, the magnetic ordering becomes short-ranged. The low-temperature increase in the ac-susceptibility associated with the SG-like cups in the underdoped crystals may also mask the inflection marking $T_N$ in the $\chi'(T)$ curves. The absence of long-ranged AFM will be confirmed by diffraction techniques\cite{jing}. Nevertheless, we show in the following that a true SG phase transition occurs in the underdoped crystals at low temperatures, which implies that any intermediate long-ranged AFM state would be frustrated, and exhibit non-equilibrium dynamical features characteristic of SG\cite{reAFM}. \revision{As illustrated in the inset of Fig.~\ref{fig-XT} for $x$=0.3}, and suggested by the comparison of the $d\chi'/dT$ data for different orientation of the magnetic field for $x$=0.4 in Fig.~\ref{fig-dXdT}, the SG-like cusp is stronger (and the overall susceptibility larger) when the magnetic field in applied perpendicularly to the $ab$-plane. 

\begin{figure}[h]
\onefigure[width=0.46\textwidth]{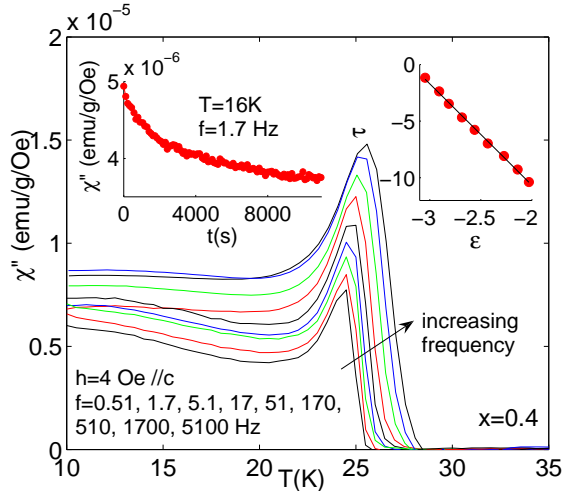}
\caption{\revision{Temperature dependence of the out-of-phase component of the ac-susceptibility $\chi''$ for the crystal with $x$=0.4 for different frequencies. The left inset illustrates the spin glass-like relaxation of  $\chi''$ with the time $t$ while the right inset shows the result of a dynamical scaling of the ($T,t$) freezing data extracted from the susceptibility curves displayed in the main frame (see footnote~\cite{note-scal}).}} 
\label{fig-SG}
\end{figure}

The  out-of-phase component $\chi''$ is negligible at all temperatures for $x$ $\geq$ 0.5. A significant $\chi''$ appears however at low temperatures for $x$ $<$ 0.5. As shown in Fig.~\ref{fig-SG}, a frequency-dependent peak is observed in the temperature range where a cusp is observed in the $\chi'-T$ curves. \revision{Below this peak, a SG-like relaxation is observed, as illustrated in the left inset. The low-frequency $\chi''$ decreases with increasing time $t$ at a constant temperature, reflecting the aging process, i.e. the rearrangement of the spin configuration toward the equilibrium state at that temperature\cite{ghost,nanopap}. A dynamical scaling analysis \cite{note-scal,nanopap} indicates that the system undergoes a SG phase transition at $T_g$ = 24K, with critical exponents similar to those of 3$D$ Ising spin glasses\cite{roland-esmo,ghost}, and a flipping time of the fluctuating entities $\tau_0$ close to that of the microscopic spin flip time (10$^{-13}$ s). Hence this SG state is three-dimensional, and nearly atomic. This means that the low-temperature magnetic state is homogeneously disordered down to the nanometer scale}. The orbital order of PCMO is rather three-dimensional\cite{jing}, which may explain the three-dimensional and isotropic character of the SG state observed in the underdoped crystals. In $R_{0.5}$Sr$_{1.5}$MnO$_4$ with two-dimensional orbital order, the long-ranged CO-OO state observed  for $R$=La is gradually replaced by a CE-glass state yielding an anisotropic, XY SG state, for $R$=Eu\cite{roland-esmo}. In the other PCMO crystals with $x$ $<$ 0.45, SG phase transitions with critical exponents similar to those of the crystal with $x$=0.4 occur, and $T_g$ increases as $x$ decreases\cite{note-expo}. In the case of $x$=0.45, only glassiness, i.e. spin-glass features such as aging, are observed, but no phase transition.\\

We summarise these findings in the electronic phase diagram presented in Fig.~\ref{fig-diag}. The phase transition temperatures $T_{\rm CO-OO}$,  $T_{\rm N}$, and $T_g$ are indicated as well as  the onsets of the short-ranged orbital ($T^*$, estimated from the ED; see Ref.~\cite{yu}) and magnetic ($T_{\rm S}^*(ab)$) correlation and the non-equilibrium dynamics ($T_f$). Three regions can be distinguished, for $x$ $<$ 0.35, 0.35 $\leq$ $x$ $<$ 0.5, and  $x$ $\geq$ 0.5. 
In the latter region, the CO-OO state is long-ranged, with a similar $T_{\rm CO-OO}$, and the AFM transition occurs at lower temperatures (see Fig.~\ref{fig-diag}). With increasing $x$ from 0.5, the CO-OO configuration depicted in the middle panel of the lower part of  Fig.~\ref{fig-diag} is rearranged to accommodate the extra holes, as observed in the optical spectra of these manganites\cite{lee} and the $x$-dependence of the modulation wave vector of the CO-OO; $q=1-x$\cite{yu} ($q$ is expressed in unit of the reciprocal lattice vector). An example of the orbital-spin configuration in that case is shown in the right panel ($x$=2/3). The accommodation of the holes within the CO-OO phase has also been observed in EuBaMn$_2$O$_6$ perovskites with no quenched disorder\cite{aka-new} as well as in single-layered systems with a fairly large quenched disorder, as La$_{1-x}$Sr$_{1+x}$MnO$_4$\cite{larochelle} ($\sigma^2$ $\sim$ 1.7$\times$10$^{-3}$ {\AA}$^2$).

In the underdoped region, $T_{\rm CO-OO}$ decreases sharply as the CO-OO state is weakened. ED studies show that the modulation wave vector of the CO-OO remains close to that of the $x$=0.5 crystal, with the commensurate value of $q$=0.5\cite{yu}, indicating that the half-doped CO-OO configuration depicted in the middle panel in the lower part of  Fig.~\ref{fig-diag} remains as extra electrons are added, in contrary to the overdoped case. Interestingly, while the orbital sector is long-ranged, the spin sector appears to remain short-ranged, as some frustration is introduced in the spin structure with these additional localised electrons locally affecting the magnetic interaction within the CE zig-zag chains (ultimately yielding the apperance of a SG state at low temperatures). In the $x$ $<$ 0.35 regions, only short-ranged orbital (CE-glass) and magnetic (SG) correlation is observed. The $x$ $\geq$ 0.5 and $x$ $<$ 0.35 regions are similar to the two regions with long-ranged, resp. short-ranged, CO-OO order in the bandwidth-disorder phase diagram of the single-layered manganites\cite{roland-diag}.

\revision{An intermediate region, with 0.35 $\leq$ $x$ $<$ 0.5, exists, where the long-ranged CO-OO is not accompanied by a long-range (AFM) magnetic state. Instead, a frustrated/short-range magnetic state is observed below $T_{\rm CO-OO}$. As shown in the above, this SG state is homogeneous down to the nanometer scale and is thus not related to the so-called macroscopic phase separation often observed in polycristalline samples or some single-crystals. Usually this macroscopic phase separation is observed when defects or impurities hinder the long-range CO-OO\cite{raveau,kimura} (the CO-OO correlation length thus becomes finite).} For example in the phase-separated Cr-doped Nd$_{0.5}$Ca$_{0.5}$MnO$_3$\cite{kimura}, Cr ions randomly replace Mn sites. Cr$^{3+}$ has the same electronic configuration as Mn$^{4+}$, with no $e_g$ electron.  Cr$^{3+}$ replacing Mn$^{3+}$ strongly affects the CO-OO as the substitution yield an immobile $e_g$ orbital deficiency\cite{kimura}. However in the present under-doping case, as Mn$^{3+}$ replaces some Mn$^{4+}$, the \revision{long-ranged} orbital order is preserved, as only a ``passive'' electron/orbital is added.  As exemplified in the left inset in the lower part of Fig.~\ref{fig-diag}, the associated super-exchange interaction locally affects the magnetic coupling yielding the magnetic frustration. In the above mentioned EuBaMn$_2$O$_6$, CO-OO rearrangements specific to these $A$-site ordered structures as well as ferromagnetism are observed. Thus the electronic phase diagram of these manganites with no quenched disorder display an asymmetry with respect to over- and under- hole doping\cite{aka-new} similar to that observed in Pr$_{1-x}$Ca$_{1+x}$MnO$_4$. Hence the relation of the order-disorder phenomena in the charge-orbital and spin sectors of manganites to the hole doping level appears rather universal.\\

To conclude, we have investigated the effect of under and over hole doping on the magnetic and charge-orbital states of Pr$_{1-x}$Ca$_{1+x}$MnO$_4$ single-crystals with weak quenched disorder. We have established the electronic phase diagram of Pr$_{1-x}$Ca$_{1+x}$MnO$_4$  as a function of the hole doping level $x$. This asymmetric phase diagram contain a region (0.35 $\leq$ $x$ $<$ 0.5) within which the long-ranged charge-orbital order is not associated with a long-ranged antiferromagnetic state, but a short-ranged magnetic arrangement exhibiting glassy features. We have discussed the origin of this frustrated magnetic state, which arises from the presence of passive $e_g$ orbitals locally favouring the antiferromagnetic super-exchange interaction instead of the ferromagnetic double exchange assuring the conductivity and ferromagnetic arrangement of each zig-zag chain of the $CE$-type structure. As for manganites with significant quenched disorder, the macroscopic phase separation stemming from the presence of local $e_g$ orbital deficiencies was not observed.


\begin{thebibliography}{0}
\bibitem{note-expo}
In the case of the $x$ = 0.4 doping, the dynamical scaling is also realized when the ac probing field parallel to the $ab$ plane, yielding similar parameters: $T_g$=24 K, $z\nu$ = 9.57, and $\tau_0$ = 1.36$\times$10$^{-13}$ s. Another single crystal with $x$ = 0.4 from a different batch prepared for soft x-ray experiments\cite{jing} was also investigated, yielding $T_g$=24 K, $z\nu$=9.35, and $\tau_0$ = 2.52$\times$10$^{-13}$ s, i.e. very similar parameters. The other dopings below $x$$<$ 0.45 also show a SG phase transition. For example for $x$=0.3, $T_g$=31 K, $z\nu$=9.87, and $\tau_0$=1.64$\times$10$^{-13}$ s. Hence for all crystals with $x$ $<$ 0.45, we obtain $z\nu$ = 9 $\pm 1$ and $\tau_0$ $\sim$ 10$^{-13 \pm 1}$ s, with $T_g$ indicated in Fig.~\ref{fig-diag}.
\end{thebibliography}
\end{document}